\documentclass[preprint,12pt]{elsarticle}

\usepackage{amssymb}
\usepackage{amsmath}
\usepackage{hyperref}
\usepackage{mdframed}

\begin{document}

\begin{frontmatter}



\title{Non-Equilibrium Dynamics in QCD and Holography}


\author{Matthias Kaminski} %% Author name
\ead{mski@ua.edu}

\affiliation{organization={Department of Physics and Astronomy, University of Alabama},%Department and Organization
            addressline={514 University Boulevard},
            city={Tuscaloosa},
            postcode={35487},
            state={AL},
            country={USA}}

%% Abstract
\begin{abstract}
The plasma generated in heavy ion collisions goes through different phases in its time evolution. While early times right after the collision are governed by far-from equilibrium dynamics, later times are believed to be well described by near-equilibrium dynamics. While the regimes of non-equilibrium are prohibitively complicated to describe within QCD, effective descriptions such as hydrodynamics provide a viable approach. In addition, holographic descriptions allow access to the full non-equilibrium dynamics at strong coupling. In this presentation, we review three examples of such hydrodynamic approaches and corresponding holographic descriptions: 1) non-equilibrium shear viscosity, 2) propagation of non-equilibrium sound waves, and 3) the non-equilibrium chiral magnetic effect.
\end{abstract}

%% Keywords

\end{frontmatter}



%%%%%%%%%%%%%%%%%%%%%%%%%%%%
\section{Introduction}
\label{sec:intro}
%
% Motivation and punchline
%
\emph{A square peg does not fit into a round hole}. This idiom illustrates what happens if an inadequate effective description (the ``round hole'') is used to fit experimental data (the ``square peg''). The task turns out to be tedious and likely yields flawed results. Therefore, we pose the question: \emph{Which effective non-equilibrium description, more concretely which formulation of hydrodynamics is adequate to describe data from heavy ion collisions, and where does this description break down?} 

% Context of topic within conference
Based on my presentation~\cite{KaminskiXQCD2025} and as a contribution to the proceedings of \emph{The 21st International Conference on QCD in Extreme Conditions~(XQCD 2025)}~\cite{XQCD2025} our topic \emph{Non-Equilibrium Dynamics} fits into the conference theme of \emph{extreme QCD}~(XQCD), because we are describing an extreme time-dependent plasma state. Holography allows computations for holographic plasma under extreme conditions, such as strong anisotropy~\cite{Chesler:2008hg,Chesler:2010bi}, strong magnetic field~\cite{Ammon:2020rvg}, large chemical potential~\cite{Erdmenger:2007ja,Kaminski:2008ai}\footnote{Note that~\cite{Kaminski:2008ai} corrects an error in the original work~\cite{Erdmenger:2007ap}.}, strong coupling~\cite{Maldacena:1997re}, and lastly but most prominently \emph{holographic plasma far from equilibrium}. These holographic models can then be used to test effective descriptions such as different formulations of hydrodynamics. 

% Overview of the problem/puzzle to be solved
%
Experimental observables in heavy ion collisions are well approximated by simulations assuming nearly perfect fluid dynamics using hydrodynamic equations already at times as early as 1fm/c~\cite{Romatschke:2017ejr,noronha-hostlerUnreasonableEffectivenessHydrodynamics2016}. 
Similarly, holographic plasma evolution displays hydrodynamic behavior at very early times~\cite{Chesler:2008hg,Chesler:2009cy,Chesler:2010bi}. 
These observations suggest that hydrodynamics (possibly in some adequately adjusted form~\cite{Romatschke:2017vte}) may be applicable in anisotropic systems with large gradients, systems which are far from global equilibrium and which may have barely reached local thermal equilibrium. 
This motivates us to study the applicability of hydrodynamics in a Bjorken-expanding holographic plasma far from equilibrium. We do so computing three quantities as functions of time: 
the shear viscosity (Sec.~\ref{sec:viscosities})~\cite{wondrak2020shear}, the speed of sound (Sec.~\ref{sec:sound})~\cite{Cartwright:2022hlg, Cartwright:2026}, and the chiral magnetic conductivity (Sec.~\ref{sec:CME})~\cite{Cartwright:2021maz}.\footnote{For lattice gauge theory considerations regarding the absence of the chiral magnetic effect in equilibrium states see the conference contribution by Garnacho et al.~\cite{GarnachoXQCD2025} and the review articles~\cite{Endrodi:2024cqn,Adhikari:2024bfa}.} 

% Summary of methods and results
%
% Details in references 
Note that in this proceeding contribution we merely summarize selected results and put them into context, while all details can be found in our original papers~\cite{Cartwright:2022hlg,wondrak2020shear, Cartwright:2021maz,Ammon:2020rvg} and the upcoming~\cite{Cartwright:2026}. 
% Methods
Nonetheless, we now provide a broad strokes overview of our calculation methods for context. In each of the three examples, we consider a gravitational setup which can be holographically mapped to $\mathcal{N}=4$ Super-Yang-Mills~(SYM) theory in some plasma state. Generalized Einstein's equations describe the time-evolution of the holographically dual spacetime metric, for which we obtain numerical solutions. These numerical data are holographically mapped to the time-dependent energy-momentum tensor and charge currents. In order to analyze these energy-momentum and current data, we systematically construct an anisotropic hydrodynamic description, beginning by composing the anisotropic constitutive equations out of all tensor structures which transform as (pseudo)scalars, (pseudo)vectors and (pseudo)tensors under the Lorentz group, subsequently restricting them by field theoretic or thermodynamic requirements~\cite{Jensen:2012jh,Jensen:2011xb,Banerjee:2012iz,Haehl:2014zda,Crossley:2015evo}.\footnote{In a formulation systematically different from ours a version of anisotropic hydrodynamics based on kinetic theory was derived and tested, which successfully extends the range of applicability beyond the range of its isotropic  version~\cite{Martinez:2010sd,Florkowski:2010cf,Ryblewski:2010bs,Ryblewski:2010tn,Ryblewski:2011aq,Ryblewski:2012rr,Florkowski:2012lba,Bazow:2013ifa,Strickland:2014pga,Alqahtani:2017mhy,Strickland:2017kux}.}

%%%%%%%%%%%%%%%%%%%%%%%%%%%%
\section{Two examples for anisotropic shear viscosities}
\label{sec:viscosities}
In order to distinguish effects caused by the anisotropy from non-equilibrium effects, we first consider our holographic plasma in an anisotropic equilibrium state in Sec.~\ref{sec:viscositiesStrongB}. In Sec.~\ref{sec:viscositiesStrongVaidya}, we consider the same holographic plasma in an isotropic non-equilibrium state. In both cases, the shear viscosity is computed as a function of time. Bjorken expansion is considered in Sec.~\ref{sec:sound} and~\ref{sec:CME}. 

%---------------------------------
\subsection{Plasma in equilibrium with anisotropy caused by strong magnetic field}
\label{sec:viscositiesStrongB}
The anisotropic equilibrium state we consider in this subsection is charged $\mathcal{N}=4$ Super-Yang-Mills plasma subject to a strong external magnetic field~\cite{Janiszewski:2015ura,Ammon:2020rvg}.\footnote{For a direct map of lattice gauge theory results to holographic results, see~\cite{Endrodi:2018ikq}. In~\cite{Endrodi:2018ikq}, a scale-invariance was revealed in the lattice gauge theory data, which allows to map the lattice data to the holographic result.} 
As an example result we highlight that the effective field theory derivation yields two distinct Kubo formulae for the shear viscosity depending on the direction of the shear. If the anisotropy direction (without loss of generality we choose the $z$-direction) is perpendicular to the plane in which the fluid is sheared (choose the $xy$-plane), then the Kubo formula for the transverse shear viscosity is given by 
\begin{equation}\label{eq:KuboLongitudinalShear}
    \eta_\perp = \lim\limits_{\omega\to 0}\frac{1}{\omega}\mathrm{Im}\langle T^{xy}\, T^{xy} \rangle(\omega, \vec{k}=0) \, ,
\end{equation}
where the transverse shear correlator in momentum space $\langle T^{xy}\, T^{xy}\rangle$ is retarded and evaluated at vanishing spatial momentum, $\vec{k}=0$ in the limit of vanishing frequency $\omega$. In contrast to this, the Kubo formula for the longitudinal shear viscosity is computed to be 
\begin{equation}\label{eq:KuboTransverseShear}
    \eta_{||} + c_\mathrm{Ohm}\,\rho_\perp + c_\mathrm{Hall}\,\tilde\rho_\perp= \lim\limits_{\omega\to 0}\frac{1}{\omega}\mathrm{Im}\langle T^{xz}\, T^{xz} \rangle(\omega, \vec{k}=0) \, ,
\end{equation}
which is different from~\eqref{eq:KuboLongitudinalShear} by the fact that now the anisotropy direction ($z$-direction) is within the plane of shear, which we chose to be the $xz$-plane, and also different by the two additional terms on the left hand side of~\eqref{eq:KuboTransverseShear}, which are related to the chiral anomaly (determining the coefficients $c_\mathrm{Ohm}$ and $c_\mathrm{Hall}$), as well as to the Ohmic resistivity in perpendicular direction, $\rho$, and the Hall resistivity in perpendicular direction, $\tilde\rho_\perp$. 
At this point, we stress that the Kubo formulae~\eqref{eq:KuboLongitudinalShear} and~\eqref{eq:KuboTransverseShear} are derived within a general effective field theory merely based on symmetries of the system and are thus valid for QCD plasma when subjected to a strong magnetic field.
\begin{figure}[hbt!] 
\centering
\includegraphics[width=0.95\textwidth]{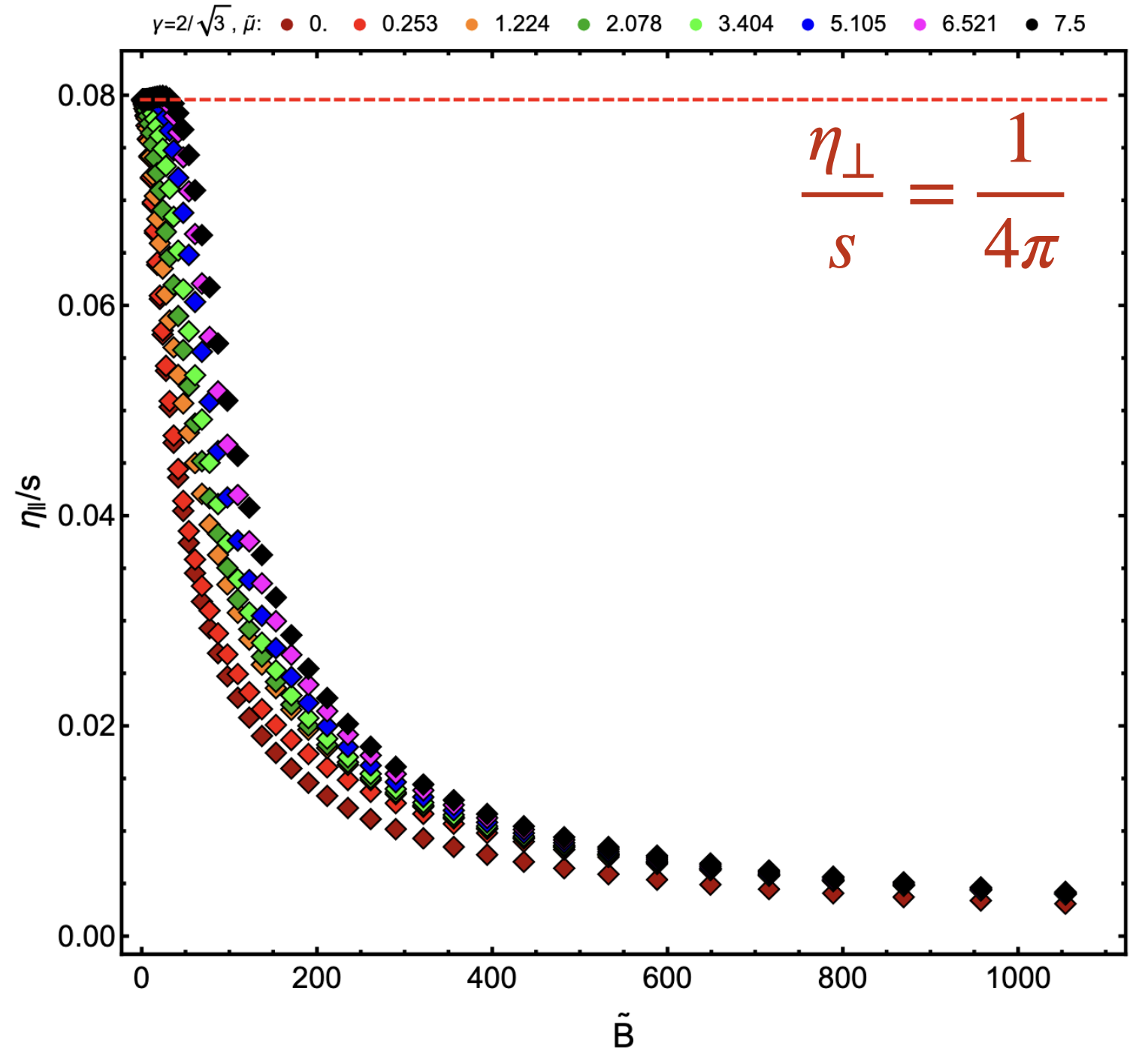}
\caption{{\it Anisotropic specific shear viscosities in equilibrium.} 
At increasing dimensionless magnetic field $\tilde B = B/T^2$ and nonzero chiral anomaly coefficient (parametrized by $\gamma=2/\sqrt{3}$), the longitudinal specific shear viscosity $\eta_{||}/s$ (colored diamonds) rapidly decreases towards large magnetic fields. Different colors correspond to the values of the dimensionless chemical potential $\mu/T$ indicated on the top. In contrast to that complex dependence, the transverse shear viscosity $\eta_\perp/s$ (dashed red line) is independent from magnetic field, temperature, chemical potential, and chiral anomaly coefficient.  
}\label{fig:AnisotropicEquilibriumShear}
\end{figure}

As collected in Fig.~\ref{fig:AnisotropicEquilibriumShear}, within $\mathcal{N}=4$ SYM theory we holographically compute qualitatively and quantitatively distinct values for the viscosities in the two directions, longitudinal along the magnetic field versus transverse to it. When normalized to the entropy density $s$, we refer to the ratio $\eta/s$ as \emph{specific shear viscosity}. 
%Already the Kubo relations are different from each other and valid for QCD plasma. 
Therefore, our results suggest that the QGP generated in heavy ion collisions can have different values for shear viscosity, depending on the alignment with the anisotropy direction. We remind the reader that the proposed Kovtun-Son-Starinets (KSS) bound $\eta/s \ge 1/(4\pi)$~\cite{Kovtun:2004de} is not a universal bound~\cite{Buchel:2008vz}, and is broken specifically for the longitudinal shear viscosity in anisotropic systems~\cite{Rebhan:2009vc,Erdmenger:2010xm,Garbiso:2020puw,Ammon:2020rvg}. 

For the shear viscosities in anisotropic systems the isotropic shear viscosity simply splits into two, as discussed above. Remarkably, there are also entirely novel transport coefficients arising, such as the \emph{shear-induced Hall conductivity}, leading to a charge current in response to a shear~\cite{Ammon:2020rvg}. A complete list of Kubo formulae for anisotropic transport coefficients in hydrodynamics subject to a strong external magnetic field is collected in tables 3, 4 and 5 of~\cite{Ammon:2020rvg}. 

% Pre-Conclusion
\begin{mdframed}
    This subsection shows that the  value of shear viscosity within a plasma depends on the direction of the external magnetic field. More generally, that magnetic field breaks rotation invariance, leading to additional terms in the constitutive relations, allowing drastically modified or entirely novel transport effects. 
\end{mdframed}

\begin{figure}[hbt!] 
\centering
\includegraphics[width=0.95\textwidth]{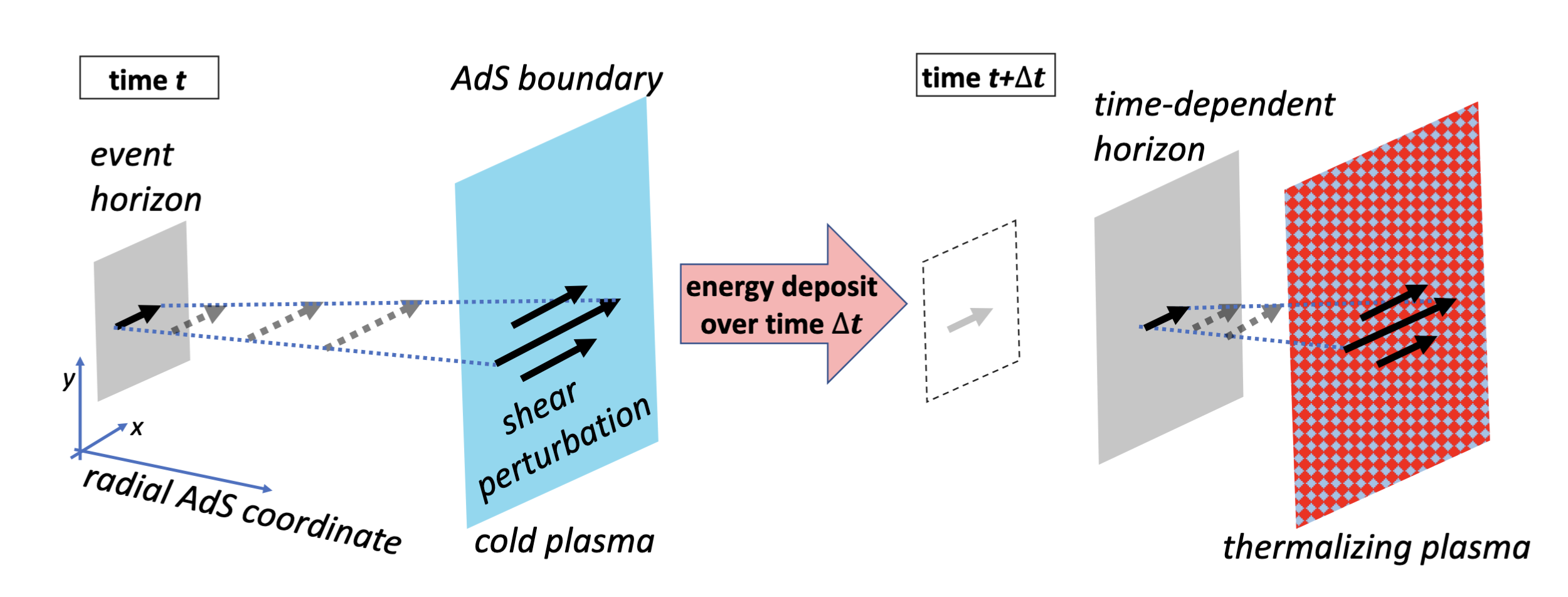}
\caption{{\it Isotropic non-equilibrium state and its holographic dual.} 
Cold plasma is heated by an amount of energy during a time $\Delta t$.   
}\label{fig:IsotropicNonequilibriumState}
\end{figure}
%
%---------------------------------
\subsection{Isotropic plasma far from equilibrium}
\label{sec:viscositiesStrongVaidya}
\begin{figure}[hbt!] 
\centering
\includegraphics[width=0.95\textwidth]{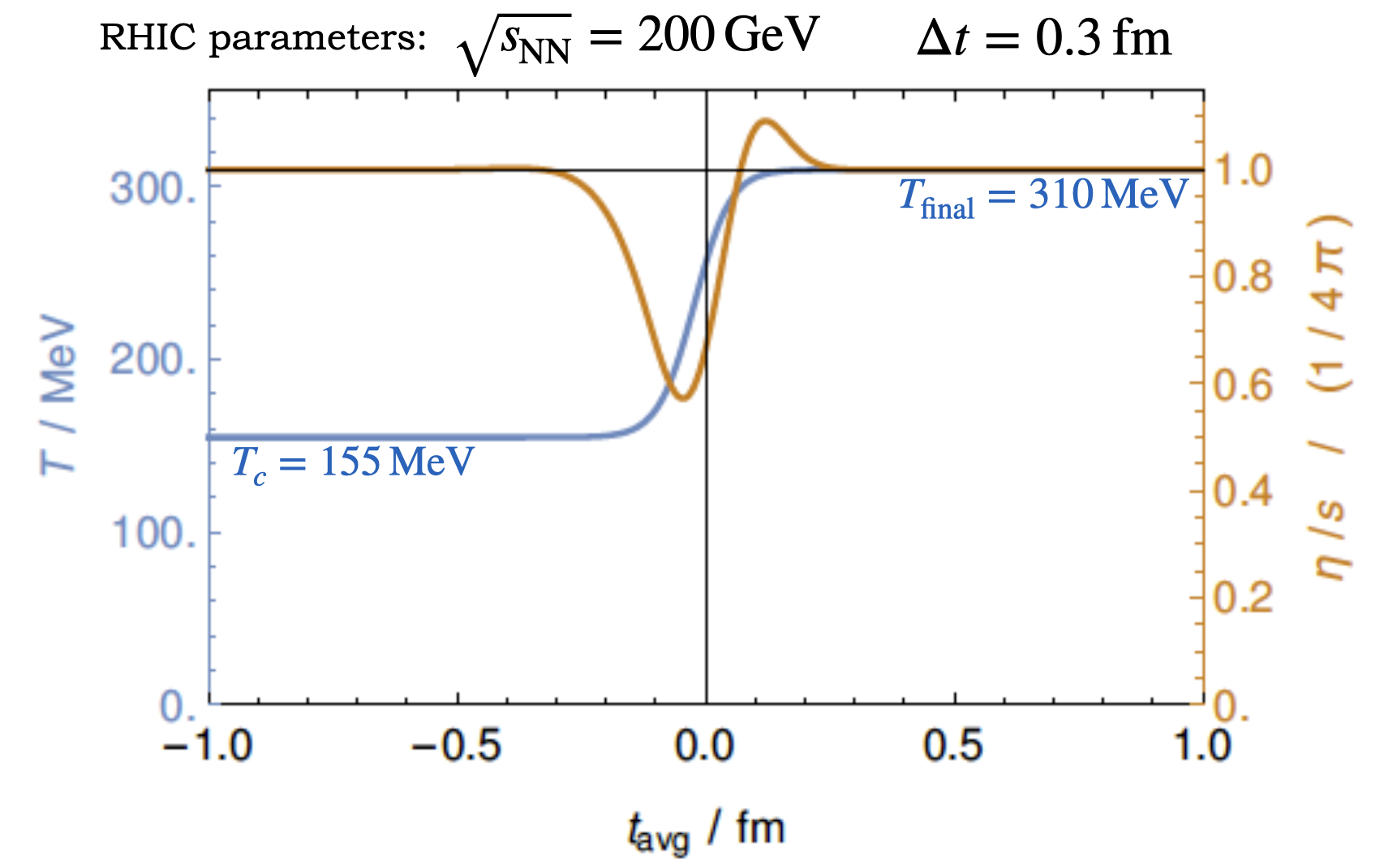}
\caption{{\it Isotropic non-equilibrium shear viscosity.} 
The specific shear viscosity normalized to the value $1/(4\pi)$ is displayed, such that a holographic isotropic equilibrium plasma would have the value $\eta/s / (1/(4\pi))=1$. }\label{fig:IsotropicNonequilibriumShear}
\end{figure}

In this subsection, the isotropic non-equilibrium state is the same SYM plasma as in Sec.~\ref{sec:viscositiesStrongB}, but now without any magnetic field. This state is created by adding an energy to a plasma, increasing its temperature from 155 MeV at time $t$ to 310 MeV at time $t+\Delta t$, see Fig.~\ref{fig:IsotropicNonequilibriumState}. This holographically corresponds to a black brane metric shifting its horizon, intuitively speaking the black hole grows and thereby increases its energy and with it its Hawking temperature. This can be achieved with the Vaidya metric, which is an analytic solution to the Einstein equations with an arbitrary time-dependent mass function~\cite{wondrak2020shear}. On top of this time-dependent Vaidya metric we introduce a shear perturbation in the metric. This is mapped to the shear correlator of the energy-momentum tensor using linear response theory and holography. 
As a result, we obtain the Kubo formula for the time-dependent shear viscosity
\begin{equation}\label{eq:KuboNonequilibriumShear}
    \eta(t_{avg}) = \lim\limits_{\omega\to 0}\frac{1}{\omega}\mathrm{Im}\langle T^{xy}\, T^{xy} \rangle(\omega, \vec{k}=0, t_{avg}) \, ,
\end{equation}
where $t_{avg}=(t_1+t_2)/2$ is the timescale on which the state changes, while the frequency $\omega$ of the shear perturbation is associated with the relative time $t_2-t_1$ through Fourier transformation, and $t_{1,2}$ are the times at which the operators are inserted in the retarded Green's function $\langle T^{xy}(t_2,\vec{x}_2)\, T^{xy}(t_1,\vec{x}_1)\rangle$. This latter correlator is first computed via holographic techniques for many combinations of values for $t_1$ and $t_2$, and then Wigner transformed to obtain $\langle T^{xy}\, T^{xy} \rangle(\omega, \vec{k}=0, t_{avg})$ which appears in the Kubo formula~\eqref{eq:KuboNonequilibriumShear}. 
As indicated in Fig.~\ref{fig:IsotropicNonequilibriumShear}, for an exemplary choices of parameters the specific shear viscosity at early average time $t_{avg}=-1 fm/c$ takes the isotropic equilibrium value 1, then decreases towards $\eta/s/(1/(4\pi))\approx 0.5$ as the temperature $T$ rises, and at later times increases to $\eta/s/(1/(4\pi))\approx 1.2$ before it drops back to 1 for all later times. 
%

% Pre-conclusion
\begin{mdframed}
In summary, this subsection teaches us that transport coefficients like the shear viscosity can change their values when evaluated in a non-equilibrium state. Together with the previous subsection, we distinguish two reasons for transport coefficients to change their values: anisotropy and non-equilibrium. There appears to be no preferred direction (increasing or decreasing) for these changes. 
\end{mdframed}

%%%%%%%%%%%%%%%%%%%%%%%%%%%%
\section{Anisotropic sound propagation in Bjorken expanding plasma}
\label{sec:sound}
\begin{figure}[hbt!] 
\centering
\includegraphics[width=0.9\textwidth]{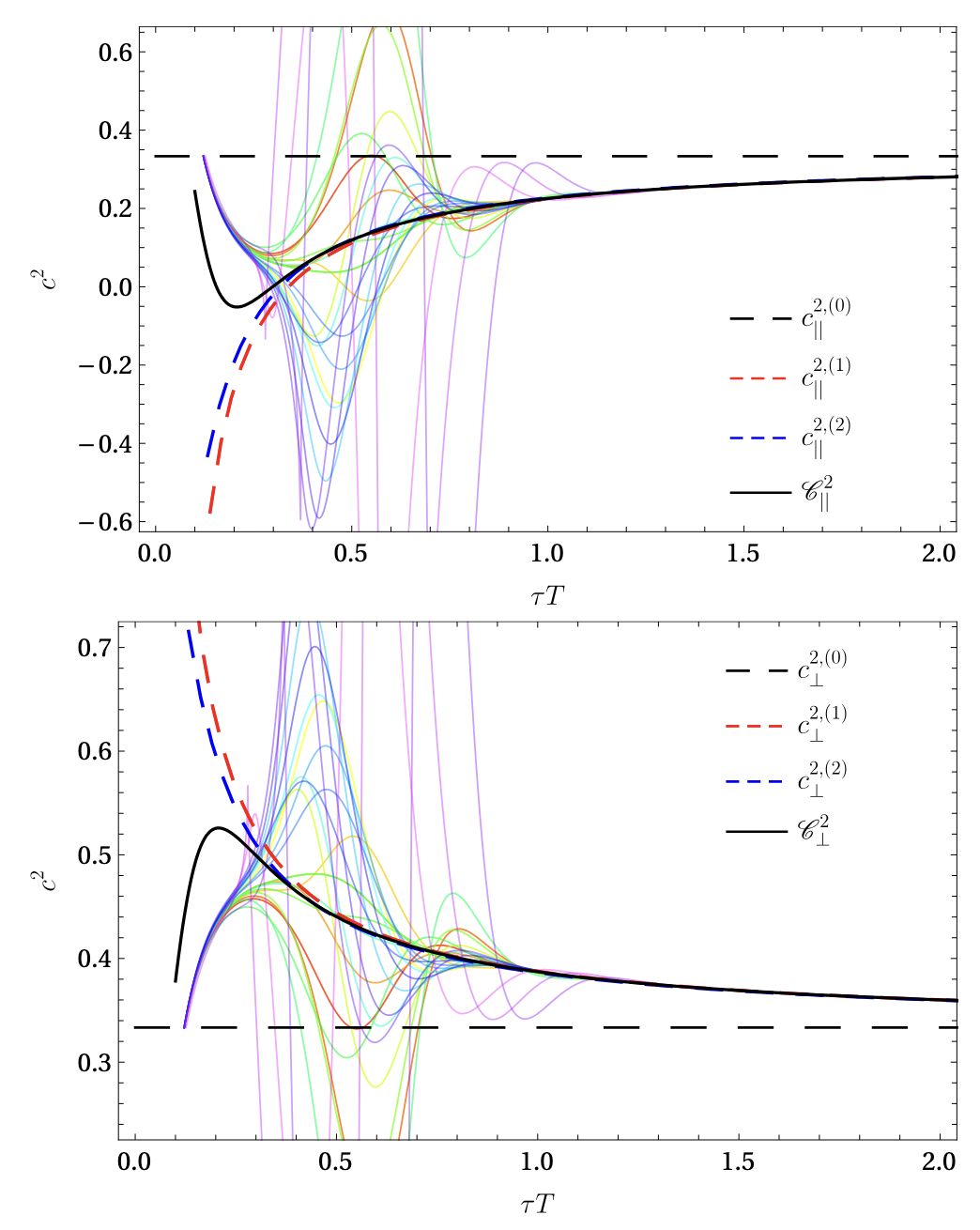}
\caption{{\it Thermodynamically defined speeds of sound in Bjorken expanding SYM plasma.} Top: Speed of sound longitudinal to the Bjorken expansion (along the anisotropy). Bottom: Transverse speed of sound. 
Thick black lines indicate the $\mathcal{N}=4$ SYM sound attractor. Dashed lines indicate zeroth, first and second order hydrodynamic approximations, see~\cite{Cartwright:2022hlg}.}\label{fig:ThermodynamicSound}
\end{figure}
In this subsection, we again consider the same holographic SYM plasma but in an uncharged anisotropic non-equilibrium state, which is dynamically anisotropic due to its longitudinal Bjorken expansion. As the time-dependent transport quantity we consider the speed of sound, $c_s$, which in isotropic equilibrium would be computed as the change of pressure with energy density at fixed entropy density $s$: $c_s^2 = \left (\partial P/\partial \epsilon\right )_s$. This is a thermodynamic equilibrium relation, which need not be valid out of equilibrium. 

As before, the anisotropy leads to two distinct speeds of sound, $c_{||}$ longitudinal and $c_\perp$ transverse to the direction of the Bjorken expansion (which is the anisotropy direction in this state). At first, we naively compute these two speeds of sound using the longitudinal and transverse pressures, $P_{||,\perp}$, naively assuming that the thermodynamic equilibrium relation can be extended to $c_{||,\perp}^2 = \left (\partial P_{||,\perp}/\partial \epsilon\right )_s$~\cite{Cartwright:2022hlg}. Results are shown in Fig.~\ref{fig:ThermodynamicSound}.  Thick solid black lines indicate the sound attractors which we computed from the known pressure anisotropy attractor of $\mathcal{N}=4$ SYM theory~\cite{spalinskiHydrodynamicAttractorYangMills2018}. Solid colored lines indicate individual time-evolutions with distinct initial conditions for the holographic Bjorken flow. In~\cite{Cartwright:2022hlg}, we already checked that thermodynamic relations such as $\epsilon +P = sT$ are not valid before proper times $\tau \approx 1/T$. This fact makes the thermodynamically defined speeds of sound even more questionable.  

In our upcoming paper~\cite{Cartwright:2026}, we improve our computation of the speeds of sound. We compute them perturbatively out of equilibrium by holographic methods and using the hydrodynamic sound dispersion relation (constructed in analogy to the isotropic case~\cite{Baier:2007ix})
\begin{equation}
    \omega(k) = \pm c_{\perp,||} \, k - i \frac{\Gamma_{\perp,||}}{2} \, k^2 
    \pm\frac{\Gamma_s}{2 c_s} \left (
    c_{\perp,||}^2 \tau^{\perp,||}_\Pi-\frac{\Gamma_{\perp,||}}{4}
    \right ) k^3
    + O(k^4) \, ,
\end{equation}    
with the sound attenuation coefficients $\Gamma_{\perp,||}$ relaxation times $\tau^{\perp,||}_\Pi$. At early times, $\tau < 1/T$, our preliminary perturbative results~\cite{Cartwright:2026} deviate quantitatively from our thermodynamically defined results, as expected. However, time evolutions of the perturbative speed of sound begins to qualitatively agree with thermodynamically defined results as early as $\tau\approx 2/T$. Details will be provided in~\cite{Cartwright:2026}. 

\begin{mdframed}
    Anisotropically expanding $\mathcal{N}=4$ SYM plasma leads to two distinct speeds of sound. Each of them follows a distinct time-evolution which depends strongly on initial conditions at early times. The difference between the two speeds of sound can be large, namely of order one. Especially at early proper times $\tau \approx 1/T$ and earlier, thermodynamically defined speeds of sound become unreliable.   
\end{mdframed}
%%%%%%%%%%%%%%%%%%%%%%%%%%%%
\section{Chiral Magnetic Effect in Bjorken expanding plasma}
\label{sec:CME}
\begin{figure}[hbt!] 
\centering
\includegraphics[width=0.9\textwidth]{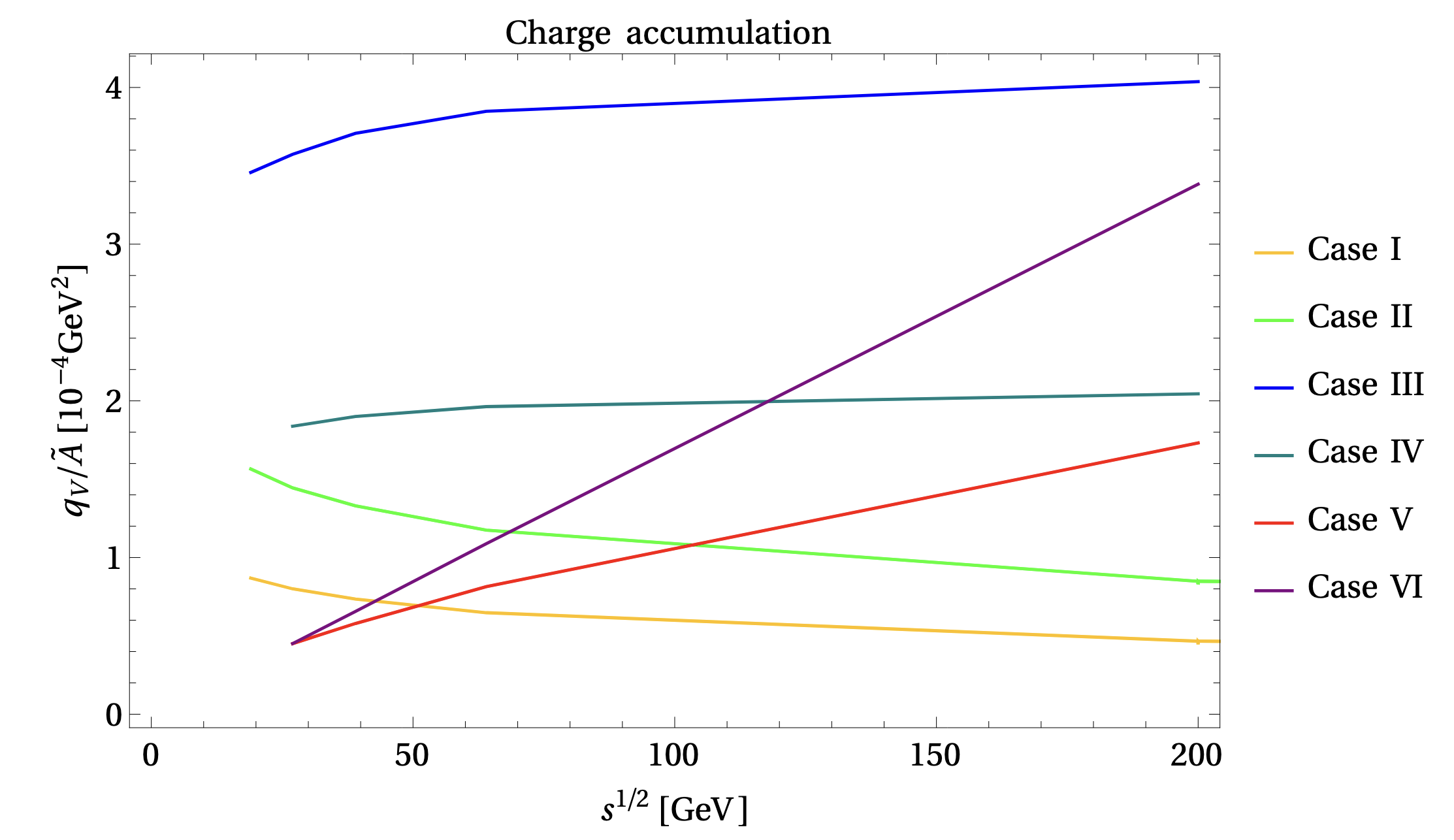}
\caption{{\it Accumulated charge due to the time-integrated chiral magnetic effect current as function of collision energy.} Different colors correspond to choice of different initial conditions for axial charge, energy density, and magnetic field (Case I-VI, see~\cite{Cartwright:2021maz}). Some cases lead to a larger charge accumulation at larger collision energy $s^{1/2}$, others lead to larger charge accumulation at smaller collision energies.} 
\label{fig:NonequilibriumCME}
\end{figure}

In this subsection, we consider a non-equilibrium $\mathcal{N}=4$ SYM plasma state which at late times follows Bjorken expansion along the “beamline”. This state is subject to a magnetic field
along one transverse direction and contains a nonzero axial charge density at vanishing electromagnetic charge density. Both, the
magnetic field and the axial charge density are time-dependent~\cite{Cartwright:2021maz}. As a result, we compute the charge which would accumulate in a detector due to the chiral magnetic effect over the duration of a collision, see Fig.~\ref{fig:NonequilibriumCME}. 
\begin{mdframed}
    If the CME is observable or not in heavy ion collisions seems to strongly depend on the details of the time evolution of the energy density, magnetic field and axial charge of the QCD plasma. 
    In our holographic model plasma, the accumulated charge depends strongly on the initial conditions (Case I-VI). However, a more recent extension of our computation favors the CME to produce the strongest signature around 10 GeV~\cite{Grieninger:2025spw}, which appears to roughly agree with the STAR experiment favoring a maximal CME at 15 GeV~\cite{XuECT2025}.
\end{mdframed}
%

%%%%%%%%%%%%%%%%%%%%%%%%%%%%
\section{Discussion}
\label{sec:discussion}
Hydrodynamics is based on symmetries. Breaking a symmetry thus has profound effects on any transport description of a given plasma. The anisotropy of the initial state generated in heavy ion collisions breaks rotation symmetry as does the anisotropic Bjorken expansion of the plasma, leading to additional terms in the hydrodynamic constitutive equations, which can lead to novel transport effects. When tested in holographic models, these novel transport effects turn out to be relevant and on the same order of magnitude as electrical conductivity and shear viscosity. 
Furthermore, known transport coefficients can significantly change their values due to non-equilibrium or due to anisotropy, as seen in the example of the shear viscosity, Sec.~\ref{sec:viscosities}.  
\begin{mdframed}
    In conclusion, QCD plasma should be analyzed using an anisotropic non-equilibrium effective description. Anisotropic hydrodynamics as defined in~\cite{Ammon:2020rvg,Garbiso:2020puw,Cartwright:2026} appears to be an adequate candidate for such a description. For example, future Bayesian analyses extending~\cite{Bernhard:2019bmu} should take this into account.  
    This conclusion is based on the effective field theory of anisotropic hydrodynamics and is thus generally valid for QCD plasma whenever large anisotropies are present or when far from equilibrium. 
\end{mdframed}

%%%%%%%%%%%%%%%%%%%%%
\paragraph*{Acknowledgments}
This work was supported, in part, by the U.S.~Department of Energy grant DE-SC0012447. 

%\appendix
%\section{Example Appendix Section}
%\label{app1}

%%       \cite{<label>} ==> [1]

%%
%Example citation, See \cite{lamport94}.

%\bibliographystyle{elsarticle-num}
%\bibliography{bib}



%\begin{thebibliography}{00}

%% For numbered reference style

%\bibitem{lamport94}
%Leslie Lamport,
%  \textit{\LaTeX: a document preparation system},
%  Addison Wesley, Massachusetts,
  %2nd edition,
  %1994.

%\end{thebibliography}
\end{document}